\documentclass[pra,aps,twocolumn,showpacs]{revtex4}
\usepackage{epsfig}
\usepackage{mathrsfs}
\usepackage{float}

\begin{document}

\title{Entanglement criteria and nonlocality for multi-mode continuous variable systems}

\author{Qingqing Sun$^{1,2}$, Hyunchul Nha$^{2,*}$, and M. Suhail Zubairy$^{1,2}$} 
\affiliation{$^1$Department of Physics and Institute of Quantum Studies, Texas A\& M University, College Station, Texas 77843, USA\\
$^2$Texas A\&M University at Qatar, Education City,
	P.O.Box 23874, Doha, Qatar}
\email{hyunchul.nha@qatar.tamu.edu}
\date{\today}

\begin{abstract}
We demonstrate how to efficiently derive a broad class of inequalities for entanglement detection in multi-mode continuous variable systems. 
The separability conditions are established from partial transposition (PT) in combination with several distinct necessary conditions for a quantum physical state, which include previously established inequalities as special cases. 
Remarkably, our method enables us to support Peres' conjecture to its full generality within the framework of Cavalcanti-Foster-Reid-Drummond multipartite Bell inequality [Phys. Rev. Lett. {\bf 99}, 210405 (2007)] that the nonlocality necessarily implies negative PT entangled states. 
\end{abstract}
%Cavalcanti-Foster-Reid-Drummond

\maketitle

\newcommand{\ds}{\displaystyle}
\newcommand{\dd}{\partial}
\newcommand{\be}{\begin{equation}}
\newcommand{\ee}{\end{equation}}
\newcommand{\beq}{\begin{eqnarray}}
\newcommand{\eeq}{\end{eqnarray}}
\newcommand{\dt}{\ds\frac{\dd}{\dd t}}
\newcommand{\dz}{\ds\frac{\dd}{\dd z}}
\newcommand{\D}{\ds\left(\frac{\dd}{\dd t} + c \frac{\dd}{\dd z}\right)}

\newcommand{\w}{\omega}
\newcommand{\W}{\Omega}
\newcommand{\g}{\gamma}
\newcommand{\G}{\Gamma}
\newcommand{\E}{\hat E}
\newcommand{\s}{\sigma}
%\newcommand{\bra}{\langle}
%\newcommand{\ket}{\rangle}
%\vspace{-0.7cm}

{\bf Introduction---}%\hspace{1cm}
%\vspace{-0.4cm}
Entanglement is the key for many applications in quantum informatics; however, there still exist a number of unresolved issues despite the progresses in the past. 
In particular, the characterization and the detection of entanglement for multipartite quantum systems are far from complete. 
Peres and Horodecki \textit{et al.} first proposed the positive partial transpose (PT) criterion as a necessary condition for separability \cite{Peres}. 
%A number of criteria using other properties of a state also exist \cite{Huang,Lee}. 
To apply this criterion in experiments for continuous variables (CVs), inequalities involving observable moments up to second-order have been derived \cite{Simon,Duan,Mancini,Werner}. 
%The violation of these provides a sufficient and necessary condition for bipartite entanglement of $1\times n$-mode Gaussian states \cite{Werner}. 
To detect bipartite entanglement for non-Gaussian CV states, new inequalities using higher order moments have been derived from the Schwarz inequalities \cite{Hillery}, uncertainty relations \cite{Agarwal,Nha, Hofmann}, and determinants of matrices \cite{Shchukin1}. These inequalities can be directly implemented in optical systems \cite{Nha,Hillery2,Shchukin3}. It was also pointed out that stronger conditions can be obtained from the ${\rm Schr\ddot{o}dinger}$-Robertson uncertainty relation (SRUR) than the Heisenberg uncertainty relation (HUR) \cite{Nha2}. 
Moreover, it was shown that the uncertainty relation approach is sufficient to detect bipartite entanglement for all negative PT states \cite{Nha3}.

For multimode entangled states, some inequalities have also been derived using similar methods \cite{Shchukin2,Li,Song, Gillet}. 
Nevertheless, our capacity for detecting and characterizing multipartite entangled states must be enhanced to a far richer level for further applications. 
In this Rapid Communication we demonstrate how to derive a broad class of inequalities for multimode CV systems following one principle: 
if a quantum state is separable, its density operator remains nonnegative under PT \cite{Peres}. 
That is, the PT density operator still represents a physical state and therefore must satisfy all necessary conditions as a quantum physical state. 
Here we particularly suggest three necessary conditions: (i) the non-negativity of $\left\langle O^{\dagger} O \right\rangle$ for any operator $O$ \cite{note}, (ii) the non-negativity of observable variances, and (iii) the satisfaction of uncertainty principle. Along these lines, we can easily obtain stronger multimode inequalities than the ones in the previous literature \cite{Hillery,Li}. %We also present new inequalities that can detect different classes of entangled states and generalize these to multimodes systems involving high-order operators. 

Remarkably, our method enables us to strongly support Peres' conjecture within the framework of Cavalcanti-Foster-Reid-Drummond (CFRD) multipartite Bell inequality \cite{Cavalcanti} 
that nonlocality necessarily implies negative PT states.
Very few results are known about the conjecture except for the most general proof 
in the $(n,2,2)$ scenario \cite{Peress}. 
Here $(n,m,o)$ refers to $n$ parties taking $m$ different measurement settings with $o$ outcomes. 
Recently, Salles {\it et al.} \cite{Salles} provided the link between nonlocality and negative PT entanglement for the first time for CVs in the $(n,2,\infty$) scenario of the CFRD inequality, but in a very restricted measurement setting, i.e. quadrature measurement only \cite{Salles}. 
In contrast, we demonstrate the full generality of the conjecture regardless of measurement settings ($o$: arbitrary) both for discrete and CVs. 
We also point out the advantages of our generalized approach, particularly of the SRUR inequality, for detecting multi-mode entangled states in comparison with the CFRD inequality. 

First we briefly introduce PT since it is the basis of all our criteria. 
Any operator $\hat{O}$ acting on a $d$-dimensional Hilbert space can generally be expanded as $\hat{O}=\sum_{i,j}O_{ij}|i\rangle\langle j|$ with basis states $|i\rangle$ ($i=1,\dots,d$).
Let us define a conjugation operator, $\hat{O}^*\equiv\sum_{i,j}O_{ij}^*|i\rangle\langle j|$. %, by taking the complex conjugate of the elements $O_{ij}$. 
Note that transposition corresponds to the conjugation operation of the Hermitian adjoint,  $\{\hat{O}^T\}_{ij}=O_{ji}=\{\hat{O}^{\dag*}\}_{ij}$.
Then, for a density operator $\rho$, one obtains the relation $
  \left\langle \hat{O} \right\rangle_{\rho^{T}}\equiv {\rm Tr}\{\hat{O}\rho^{T}\}=
{\rm Tr}\{\rho\hat{O}^T\}= \left\langle \hat{O}^{\dag*} \right\rangle_{\rho}$, for trace is preserved by transposition. 
It can be readily extended to PT as
\begin{eqnarray}
 \left\langle \hat{O}_A\hat{O}_B \right\rangle_{\rho^{PT}}=\left\langle \hat{O}_A\hat{O}_B^{\dag*} \right\rangle_{\rho},
 \label{eqn:PT-rule1}
\end{eqnarray}
where a whole system is split to two subsystems $\{A,B\}$ with transposition on $B$, and ${\hat O}_{A,B}$ is the operator acting on $A$ and $B$, respectively. 
Therefore, the quantum average over $\rho^{PT}$ can always be expressed 
by an experimentally accessible average through Eq.~(\ref{eqn:PT-rule1}). 
For instance, in a two-mode system represented by the boson operators $a$ and $b$, we immediately obtain \cite{Nha} 
\beq
	\left\langle a^{\dagger k} a^{l} b^{\dagger p} b^{q} \right\rangle_{\rho^{PT}} = \left\langle a^{\dagger k} a^{l} b^{\dagger q} b^{p} \right\rangle_{\rho},
\label{eqn:PT-rule}	
\eeq
for $b^{\dagger p} b^{q}$ has only real elements in number-state basis. 

{\bf Two Modes---} 
%\vspace{-0.2cm}
We first illustrate using two-mode CV systems how the three physical properties required as a quantum state can be employed to derive entanglement conditions. %Let $a$ and $a^{\dagger}$ be the annihilation and the creation operator of the first mode, and $b$ and $b^{\dagger}$ of the second. 
Let us first define an operator $O = r (a - \left\langle a \right\rangle) + \frac{1}{r} (b - \left\langle b \right\rangle)$ and the EPR-like operators
\beq
  u = r x_{a} + \frac{1}{r} x_{b},    ~~v = r p_{a} - \frac{1}{r} p_{b},
\eeq
where $x_{a} = (a^{\dagger} + a)/\sqrt{2}, p_{a} = i(a^{\dagger} - a)/\sqrt{2}$, and similarly for mode $b$ ($r$: nonzero real). 
Under PT on mode $b$, the expectation value of the positive operator $O^{\dagger} O$ becomes 
\beq
\left\langle O^{\dagger} O \right\rangle_{\rho^{PT}} = \frac{1}{2} \left[ \left\langle (\Delta u)^{2} \right\rangle_{\rho} + \left\langle (\Delta v)^{2} \right\rangle_{\rho} - r^{2} - \frac{1}{r^{2}} \right].
\eeq 
If the original density operator $\rho$ is separable, it still remains nonnegative under PT. 
Therefore, the quantum average of a positive operator must also be nonnegative, i.e., $\left\langle O^{\dagger} O \right\rangle_{\rho^{PT}} \geq 0$, 
which immediately gives the separability condition in Eq.~(3) of Ref. \cite{Duan}.

Next, let us consider the operators
\beq
J_{x}&=&\frac{1}{2}(a^{\dagger}b + ab^{\dagger}), ~~ K_{x}=\frac{1}{2}(a^{\dagger}b^{\dagger} + ab),  \nonumber \\
J_{y}&=&\frac{1}{2i}(a^{\dagger}b - ab^{\dagger}),  ~K_{y}=\frac{1}{2i}(a^{\dagger}b^{\dagger} - ab),  \nonumber \\
J_{z}&=&\frac{1}{2}(N_{a} - N_{b}),  ~~~ K_{z}=\frac{1}{2}(N_{a} + N_{b} + 1),
\label{eq:J}
\eeq
where $N_{a}=a^{\dagger}a$ and $N_{b}=b^{\dagger}b$ are the number operators. 
Due to the commutation relations, $J_{i}~(i=x,y,z)$ forms the SU(2) algebra, and $K_{i}$ forms the SU(1,1) algebra. 
Under PT on the second mode, the variances are obtained using the rule in Eq.~(\ref{eqn:PT-rule}) as
\beq
  (\Delta J_{i})^{2}_{\rho^{PT}} = (\Delta K_{i})^{2}_{\rho} - \frac{1}{4}~~~~~(i=x,y).   
 \label{eq:Jx-variance}
\eeq
Therefore, the separability conditions follow from the positivity of the variances $(\Delta J_{i})^{2}_{\rho^{PT}}$ $(i=x,y)$ as
\beq
  (\Delta K_{i})^{2}_{\rho} \geq \frac{1}{4} ~~~~~(i=x,y).
\label{Eq:Hillery}
\eeq
A violation of these inequalities provides a sufficient condition for inseparablity and  
the same conditions were obtained using a different method in Ref. \cite{Hillery}.

The third property that must be satisfied by a physical state is the HUR. 
From the commutation relation $[J_{x},J_{y}] = iJ_{z}$, we have $(\Delta J_{x})^{2}_{\rho^{PT}} (\Delta J_{y})^{2}_{\rho^{PT}} \geq \frac{1}{4} \left| \left\langle J_{z} \right\rangle_{\rho^{PT}} \right|^{2}$. 
On substituting Eqs.~(\ref{eq:J}) and (\ref{eq:Jx-variance}), we obtain a necessary condition for separability
\beq
  \left[ (\Delta K_{x})^{2}_{\rho} - \frac{1}{4} \right] \left[ (\Delta K_{y})^{2}_{\rho} - \frac{1}{4} \right] \geq \frac{1}{16} \left| \left\langle N_{a} - N_{b} \right\rangle_{\rho} \right|^{2}.
\label{Eq:a1b1}
\eeq
The above condition is stronger than the one in Eq.~(\ref{Eq:Hillery}): the right hand side (RHS) of Eq.~(\ref{Eq:a1b1}) is non-negative  and $(\Delta K_{x})^{2}_{\rho}$ and $(\Delta K_{y})^{2}_{\rho}$ cannot be both smaller than $1/4$. If both of them were less than $1/4$, it would violate the regular HUR of $[K_{x},K_{y}] = - iK_{z}$, i.e.,
\beq
   (\Delta K_{x})^{2}_{\rho} + (\Delta K_{y})^{2}_{\rho}  \geq 2 (\Delta K_{x})_{\rho} (\Delta K_{y})_{\rho} \geq  \left| \left\langle K_{z} \right\rangle_{\rho} \right| \geq \frac{1}{2}, 
\eeq
which must be satisfied by all states, separable or not.

The HURs from other commutation relations can also be useful. 
For example, from the HUR of $[K_{y},K_{z}] = iK_{x}$, we obtain the necessary condition for separability as
\beq
  \left[ (\Delta J_{y})^{2}_{\rho} + \frac{1}{4} \right] (\Delta K_{z})^{2}_{\rho} \geq \frac{1}{4} \left| \left\langle J_{x} \right\rangle_{\rho} \right|^{2}.
\label{Eq:ex}
\eeq
As an example, for the class of states with total excitation fixed, e.g. $( \left| i \right\rangle_{a} \left| j+1 \right\rangle_{b} + \left| i+1 \right\rangle_{a} \left| j \right\rangle_{b}  ) / \sqrt{2}$ ($i,j=0,1,\cdots$), we have $ (\Delta K_{z})^{2}_{\rho} = 0 $ on the left side of Eq.~(\ref{Eq:ex}). 
On the other hand, the average $ \left\langle J_{x} \right\rangle_{\rho} $ is nonzero, therefore the inequality is violated.

The above ideas can also be extended to higher order terms. For instance, let us define the $L$ operators
\beq
L_{x}&=&\frac{1}{2}(a^{\dagger m}b^{n} + a^{m}b^{\dagger n}), \nonumber\\
L_{y}&=&\frac{1}{2i}(a^{\dagger m}b^{n} - a^{m}b^{\dagger n}),   \nonumber \\
L_{z}&=&\frac{1}{2}[a^{\dagger m}b^{n},a^{m}b^{\dagger n}],
\eeq
where $m$ and $n$ are positive integers. 
Similarly, we define the $H_i$ operators only by exchanging $b\leftrightarrow b^\dag$ in $L_i$ ($i=x,y,z$).
%\beq
%L_{x}&=&\frac{1}{2}(P_{mn} + P_{mn}^\dag),~L_{y}=\frac{1}{2i}(P_{mn} - P_{mn}^\dag),   \nonumber \\
%L_{z}&=&\frac{1}{2}(P_{mn} P_{mn}^\dag - P_{mn}^\dag  P_{mn}),
%\eeq
%and
%\beq
%H_{x}&=&\frac{1}{2}(Q_{mn} + Q_{mn}^\dag), ~ H_{y}=\frac{1}{2i}(Q_{mn}-Q_{mn}^\dag),   \nonumber \\
%H_{z}&=&\frac{1}{2}(Q_{mn} Q_{mn}^\dag-Q_{mn}^\dag Q_{mn}),
%\eeq
%where $P_{mn}\equiv a^{\dag m}b^{n}$, $Q_{mn}\equiv a^{m}b^{n}$,  and $\{m,n\}$ are positive integers. 
These operators do not form any of previously known algebra; however, we are only concerned with their commutation relations which provide the uncertainty relations.
The variances of a state under PT on the second mode are
\beq
  (\Delta L_{i})^{2}_{\rho^{PT}} &=& (\Delta H_{i})^{2}_{\rho} - \left\langle N_{mn} \right\rangle_{\rho} ~~~~~(i=x,y),
\label{eq:highLvariance}
\eeq
where $N_{mn}\equiv\frac{1}{4}[a^{m},a^{\dagger m}]\otimes [b^{n},b^{\dag n}]$. 
On substituting Eq.~(\ref{eq:highLvariance}) into the HUR of $\left[ L_{x}, L_{y} \right] = i L_{z}$ under PT, we obtain a separability condition as
\beq
  \left[ (\Delta H_{x})^{2}_{\rho} - \left\langle N_{mn} \right\rangle_{\rho} \right] \left[ (\Delta H_{y})^{2}_{\rho} - \left\langle N_{mn} \right\rangle_{\rho} \right] \geq \frac{1}{4} \left| \left\langle L_{z} \right\rangle_{\rho} \right|^{2}.
\eeq
The inequality in Eq.~(\ref{Eq:a1b1}) is just a special case of $m=n=1$. 
On the other hand, if we define the operator $N_{+} = (na^{\dagger}a + mb^{\dagger}b)/2mn$, it provides commutation relations
$\left[ H_{x}, N_{+} \right] = -i H_{y}, ~~  \left[ H_{y}, N_{+} \right] = i H_{x}$,
and the HUR under PT reads
\beq
  \left[ (\Delta L_{y})^{2}_{\rho} + \left\langle N_{mn} \right\rangle_{\rho} \right] (\Delta N_{+})^{2}_{\rho} \geq \frac{1}{4} \left| \left\langle L_{x} \right\rangle_{\rho} \right|^{2},
\eeq
For a class of states $( \left| i \right\rangle_{a} \left| j+n \right\rangle_{b} + \left| i+m \right\rangle_{a} \left| j \right\rangle_{b}  )/\sqrt{2}$, we have $ (\Delta N_{+})^{2}_{\rho} = 0 $ but $ \left| \left\langle L_{x} \right\rangle_{\rho} \right| > 0 $. 
The inequality is thus violated, which verifies non-Gaussian entanglement. 

%\vspace{-0.4cm}
{\bf Multimode Entanglement---}
 Now let us extend further to multimode systems with the same ideas outlined above. 
We first define $n$-mode operators as $L_{x} =\frac{1}{2} (M_n+M_n^\dag)$, $L_{y} = \frac{1}{2i}(M_n-M_n^\dag),$ and  
 $L_{z} = \frac{1}{2} [M_n,M_n^\dag]$, and construct the operator $M_n$ in the most-general normally-ordered form. 
That is, $M_n\equiv \prod_{k=1}^n a^{\dagger S'_{k}}_{k} a^{S_{k}}_{k}$, where the integers $S'_{k}$ and $S_{k}$ are the powers of the {\it k}th mode operators. 
Without loss of generality, PT can be taken on the modes $m+1$ to $n$, and we then define operators $H_x$ and $H_y$ 
by letting $S_{k}'\leftrightarrow S_k$ ($k=m+1,\cdots,n$) in $L_{x}$ and $L_{y}$, respectively. 
%where $M\equiv a^{\dagger S_{1}}_{1} \cdots a^{\dagger S_{m}}_{m} a^{S_{m+1}}_{m+1} \cdots a^{S_{n}}_{n}$, and the integer $S_{i}$ is the power of the $i$-th mode operator.
%If PT is taken for the mode $m-k+1$ to $m+l$ $ (k \leq m, l \leq n-m)$, then we define new operators $H_x$ and $H_y$ 
%by interchanging $a_i\leftrightarrow a_i^\dag$ ($i=m-k+1,\cdots,m+l$) from $L_{x}$ and $L_{y}$, respectively. 
The relations between the variances are then given by 
\beq
  (\Delta L_{i})^{2}_{\rho^{PT}} = (\Delta H_{i})^{2}_{\rho} - \left\langle N \right\rangle_{\rho}~~~~~(i=x,y), 
\label{eq:multiLvariance}
\eeq
where the operator $N$ is given by 
\begin{eqnarray}
N\equiv\frac{1}{4}\left[M_1,M_1^\dag\right]\otimes\left[M_2,M_2^\dag\right].
\label{eqn:nop}
\end{eqnarray} 
Here $M_1\equiv\prod^{m}_{i=1} a^{\dagger S'_{i}}_{i} a^{S_{i}}_{i}$ refers to the untouched modes and 
$M_2\equiv\prod^{n}_{j=m+1} a^{\dagger S_{j}}_{j} a^{S'_{j}}_{j}$ to the modes under transposition.

The HUR of $\left[ L_{x}, L_{y} \right] = i L_{z}$ under PT together with Eq.~(\ref{eq:multiLvariance}) leads to the separable condition
\beq
  \left[ (\Delta H_{x})^{2}_{\rho} - \left\langle N \right\rangle_{\rho} \right] \left[ (\Delta H_{y})^{2}_{\rho} - \left\langle N \right\rangle_{\rho} \right] \geq \frac{1}{4} \left| \left\langle L_{z} \right\rangle_{\rho} \right|^{2}.
\label{Eq:generalinequ}
\eeq
The usefulness of this generalization will be discussed later. 
Now we illustrate the power of our approach in addressing multipartite entangled systems particularly by providing in the most general form the connection between nonlocality in the framework of CFRD  inequality \cite{Cavalcanti} and NPT entanglement. 

%\vspace{-0.5cm}
{\bf Nonlocality in Multimode Systems---}
%\vspace{-0.2cm}
We first consider the CFRD Bell inequality for a general multipartite system. 
The inequality was derived based on the nonnegativity of observable variances using non-contextual hidden variables, i.e., by ignoring the noncommutativity of quantum operators. 
Defining the operator $Z_k\equiv X_k+iY_k$ with two local observables $X_k$ and $Y_k$ at {\it k}th mode and $C_n={\tilde X}_n+i{\tilde Y}_n\equiv\prod_{k=1}^n(X_k+iY_k)$, 
the Bell inequality reads $\langle{\tilde X}_n\rangle^2+\langle{\tilde Y}_n\rangle^2 \leq \left<\prod_{k=1}^n(X_k^2+Y_k^2)\right>$ \cite{Cavalcanti}, which can be written as
\begin{eqnarray}
\left|\left<\prod_{k=1}^nZ_k\right>\right|^2\leq\frac{1}{2^n}\left<\prod_{k=1}^n(Z_k^\dag Z_k+Z_kZ_k^\dag)\right>.
\label{eqn:CFRD}
\end{eqnarray}
%In this Letter, we consider a broad class of local observables such that the difference between $Z_kZ_k^\dag$ and $Z_k^\dag Z_k$ is a positive operator, i.e., $Z_kZ_k^\dag-Z_k^\dag Z_k=\pm\pi_k$, ($\pi_k$: positive).  
For comparison, let us now turn attention to a regular HUR in sum-form, $(\Delta L_1)^2+(\Delta L_2)^2\ge|\langle[L_1,L_2]\rangle|$, for two Hermitian operators $L_1=P+P^\dag$ and $L_2=i(P-P^\dag)$, where $P$ is an arbitrary operator acting on a multipartite system. 
The HUR is then reduced to 
\begin{eqnarray}
|\langle P\rangle|^2\leq {\rm min}\{\langle P^\dag P\rangle\ , \langle PP^\dag\rangle\}.
\label{eqn:HURRR}
\end{eqnarray}
If one particularly takes $P=\prod_{k=1}^nZ_k$ in the above, the HUR then reads 
\begin{eqnarray}
\left|\left<\prod_{k=1}^nZ_k\right>\right|^2\leq{\rm min}\{ \left<\prod_{k=1}^nZ_k^\dag Z_k\right> , \left<\prod_{k=1}^nZ_kZ_k^\dag \right>\}.
\label{eqn:HURR}
\end{eqnarray}
%Now one can see that the violation of CFRD inequality does not occur if all the local observables satisfy $Z_kZ_k^\dag>Z_k^\dag Z_k$: 
%The RHS of Eq.~(\ref{eqn:HURR}), which must be satisfied by any physical states, is smaller than that of Eq.~(\ref{eqn:CFRD}). 
%The same argument applies to the case of 
%$Z_kZ_k^\dag<Z_k^\dag Z_k$. 
%Therefore, without loss of generality, we must choose the observables as  $Z_kZ_k^\dag-Z_k^\dag Z_k=\pi_k$ for one set of modes $\{k=1,\dots,j\}$ and $Z_kZ_k^\dag-Z_k^\dag Z_k=-\pi_k$ 
%for the other set $\{k=j+1,\dots,n\}$ to reveal nonlocality. 
On expanding the terms in the RHS of the CFRD inequality~(\ref{eqn:CFRD}), one can see that it is larger than, or equal to, the minimum over all possible values of $\left<\prod_{k\in {\cal N}}Z_k^\dag Z_k\prod_{k\in {\cal A}}Z_kZ_k^\dag \right>$. 
Here, $n$ modes can be divided to two disjoint groups ${\cal N}\cup {\cal A}=\{1,\dots,n\}$ in $2^n$ different ways, where the group ${\cal N}$ refers to the modes with the "normal"-ordered operator $Z_k^\dag Z_k$ 
and the group ${\cal A}$ to the modes with the "antinormal"-ordered operator $Z_kZ_k^\dag$.
Let us denote the two groups corresponding to the minimum by ${\cal N}_m$ and ${\cal A}_m$. 
Comparing Eqs.~(\ref{eqn:CFRD}) and~(\ref{eqn:HURR}), one can see that the violation of the CFRD inequality does not occur for the cases of ${\cal N}_m=\phi$ or ${\cal A}_m=\phi$ ($\phi$: null-set), for which the RHS of Eq.~(\ref{eqn:CFRD}) is not smaller than that of Eq.~(\ref{eqn:HURR}). 
Note that the HUR in Eq.~(\ref{eqn:HURR}) must be satisfied by any quantum states. Therefore, we rule out the cases of ${\cal N}_m=\phi$ or ${\cal A}_m=\phi$ in the following. 

We now show that a PT separability condition is necessarily violated if the CFRD inequality~(\ref{eqn:CFRD}) is violated. 
Let $P=\prod_{k\in {\cal N}_m }Z_k\prod_{k\in {\cal A}_m}Z_k^{\dag*}$ in the general HUR of Eq.~(\ref{eqn:HURRR}) and take transposition on the subset of modes $k\in {\cal A}_m$. 
Due to the rule in Eq.~(\ref{eqn:PT-rule1}), the separability condition emerges as 
\begin{eqnarray}
\left|\left<\prod_{k=1}^nZ_k\right>\right|^2\leq \left<\prod_{k\in {\cal N}_m}Z_k^\dag Z_k\prod_{k\in {\cal A}_m}Z_kZ_k^\dag\right>.
\label{eqn:sepp}
\end{eqnarray}
Since the RHS of Eq.~(\ref{eqn:CFRD}) is not smaller than that of Eq.~(\ref{eqn:sepp}), Peres' conjecture is supported, 
that is, the CFRD inequality is violated only by an NPT multipartite entangled state. 

Note that the above proof is valid for any two local observables $X_k$ and $Y_k$ at each party regardless of Hilbert-space dimension and the number of measurement outcomes. 
In this Rapid Communication, we illustrate the significance of our general approach to higher-orders by considering a broad class of local observables for CVs, namely, $Z_k\equiv a_k^{m_k}e^{-i\theta_k}$ ($k=1,\dots,j$) and $Z_k\equiv a_k^{\dag m_k}e^{i\theta_k}$ ($k=j+1,\dots,n$), where  $m_k$ is a positive integer. 
The support of Peres' conjecture by Salles {\it et al.} refers to a limited case of $m_k=1$ for each mode, which corresponds to quadrature measurement \cite{Salles}. 
Cavalcanti {\it et al.} investigated the violation of the Bell inequality also for $m_k=1$ \cite{Cavalcanti}. With the generalization to higher-orders, however, one can demonstrate nonlocality to a richer level. Consider, for example, the state $\left| \Psi \right\rangle = c_{0} \left| 0_{\{1\}},\ldots,0_{\{\frac{n}{2}\}},1_{\{\frac{n}{2}+1\}},\ldots,1_{\{n\}} \right\rangle + c_{1} \left| 2_{\{1\}},1_{\{2\}},\ldots,1_{\{\frac{n}{2}\}},0_{\{\frac{n}{2}+1\}},\ldots,0_{\{n\}} \right\rangle$, where the notation $N_{\{j\}}$ refers to $N$ photons in the mode $j$.  
The CFRD inequality is not at all violated with the choice of $m_k=1$, but it is so with the modification to $m_{1} =2$ for the number of modes $n \geq 14$.  

The separability conditions in Eq.~(\ref{eqn:sepp}) are obviously useful on their own. 
Note that the inequality~(\ref{eqn:sepp}) is valid as a separability condition for any pair of groups $\{{\cal N},{\cal A}\}$, not necessarily corresponding to minimum $\{{\cal N}_m,{\cal A}_m\}$. 
There are certain multimode states that violate the inequality (\ref{eqn:sepp}) but not the Bell inequality (\ref{eqn:CFRD}). 
As an example, let us consider the multimode N00N state, $\left| \Psi \right\rangle = c_{0} \left|N_{\{1,\dots,\frac{n}{2}\}},0_{\{\frac{n}{2}+1,\dots,n\}}\right\rangle + c_1 \left|0_{\{1,\dots,\frac{n}{2}\}},N_{\{\frac{n}{2}+1,\dots,n\}}\right\rangle$, where the first $\frac{n}{2}$ modes or the second $\frac{n}{2}$ modes have $N$ photons.
Let $Z_k\equiv a_k^{m_k}e^{-i\theta_k}$ ($k=1,\dots,\frac{n}{2}$) and $Z_k\equiv a_k^{\dag m_k}e^{i\theta_k}$ ($k=\frac{n}{2}+1,\dots,n$) with all $m_{k} = N$. 
Inequality~(\ref{eqn:sepp}), which becomes $\left|\left<\prod_{k=1}^{\frac{n}{2}}a_k^N\prod_{k=\frac{n}{2}+1}^{n}a_k^{\dag N}\right>\right|^2\leq \left<\prod_{k=1}^na_k^{\dag N} a_k^N\right>$, is then violated for any number of modes $n$, whereas the CFRD inequality is satisfied.

In general, some practical advantages can be achieved if one uses the product form (PF) of the HUR, $\Delta L_1\Delta L_2\ge\frac{1}{2}|\langle[L_1,L_2]\rangle|$, instead of the sum form (SF) $(\Delta L_1)^2+(\Delta L_2)^2\ge|\langle[L_1,L_2]\rangle|$. 
It is because of the relation $(\Delta L_1)^2+(\Delta L_2)^2\ge2\Delta L_1\Delta L_2$ when $\Delta L_i$ ($i=1,2$) is nonnegative: if the SF is violated under PT, so is the PF, but the converse is not true in general. If $\Delta L_i$ becomes negative under PT, it is a signature of entanglement as such, so we exclude those cases in this discussion.
For instance, in the above case of N00N-state, the product-form-based inequality under PT, which belongs to the class of separability conditions in Eq.~(\ref{Eq:generalinequ}), is violated in the much lower order of $m_{k} = \frac{N}{2}$ at each party on the condition ${\rm Re}\{c_0c_1^*\prod_{k=1}^{n/2}e^{-2i(\theta_k-\theta_{k+n/2})}\}\ne0$, whereas the sum-form-based one in Eq.~(\ref{eqn:sepp}) is so only in the order $m_{k} = N$. 
Moreover, a further advantage follows if the SRUR, $(\Delta L_1)^2(\Delta L_2)^2\ge\frac{1}{4}|\langle[L_1,L_2]\rangle|^2+\frac{1}{4}\langle\Delta L_1 \Delta L_2+\Delta L_2 \Delta L_1\rangle^2$, where an additional nonnegative term appears in the right, is used instead of any forms of the HUR. 
The SRUR is insensitive to local phase errors in the prepared entangled state, since it is invariant under local phase operations. Thus, the separability inequality is unconditionally violated at $m_{k} = \frac{N}{2}$.  
In general, the set of detectable multipartite entangled states can be characterized by the inclusion map, $\{\rm CFRD\}\subset  \{\rm SF~of~HUR\} \subset \{\rm PF~of~HUR\} \subset \{\rm SRUR\}$.

In summary, we have demonstrated how to efficiently derive separable conditions for general multimode CV systems, which include all CV inequalities previously obtained as special cases. 
The physical quantities in these inequalities can be experimentally measured via homodyne correlation techniques \cite{Shchukin3}. 
The principle we follow here is simple yet powerful: under PT, any separable density operator should still describe a physical state and therefore satisfy all necessary conditions as a quantum state. Although we specifically employed three necessary conditions, % of non-negative $\left\langle O^{\dagger} O \right\rangle$, non-negative variances, and HUR (SRUR), 
other general conditions should enable one to induce new inequalities as well. 

We also showed the validity of Peres' conjecture in the framework of CFRD inequality that nonlocality necessarily implies NPT entanglement regardless of local observables both for discrete and CVs. 
In future, we may need to further prove or disprove Peres' conjecture beyond the CFRD inequality as its violation is not a necessary condition, though sufficient, to reveal nonlocality. 

This work is supported by the Qatar National Research Fund (QNRF). M. S. Z. would like to thank Mark Hillery for helpful discussions.
%*email: hyunchul.nha@qatar.tamu.edu
%\vspace{-0.5cm}
%%%%%%%%%%%%%%%%%%%%%%%%%%%%%%%%%%%%%%%%%%%%%%%%%%%%%%%%%%%%%


\begin{thebibliography}{99}
\bibitem{note} The satisfaction of $\rm Tr \{O^\dag O\rho\}\ge0$ for arbitrary operator $O$ is in fact the definition of positivity of $\rho$. 
What we suggest here is that by directly employing certain operators $O$, rather than establishing hierarchy in an indirect way taken in \cite{Shchukin1}, 
one can derive useful entanglement criteria.
\end{thebibliography}
\end{document}